# Data Encryption based on 7D Complex Chaotic System with Cubic Memristor for Smart Grid

**Lei Kou** [1], **Zhe Huang** [1], **Cuimei Jiang** [1,*], **Fangfang Zhang** [1,*], **Wende Ke** [2], **Junhe Wan** [1], **Hailin Liu** [1], **Hui Li** [1]

1 *Qilu University of Technology (Shandong Academy of Sciences), Jinan, China*
2 *Department of Mechanical and Energy Engineering, Southern University of Science and Technology, Shenzhen, China*

Correspondence*:
Cuimei Jiang,
Jiangcuimei2004@163.com
Fangfang Zhang,
zhff4u@163.com

**ABSTRACT**

The information security has an irreplaceable position in the smart grid (SG). In order to avoid the malicious attack and ensure the information security, the cryptographic techniques are essential. This paper focuses on the encryption techniques to ensure the information security of SG. Firstly, an unusual 7-dimensional complex chaotic system (7D-CCS) combined with the cubic memristor is introduced. Besides its phase portraits, Lyapunov exponent, 0-1 test, complexity, and bifurcation diagram are investigated. Then, with the proposed 7D-CCS, we design a data encryption algorithm to ensure the encryption security. Finally, the data and monitoring images in SG are encrypted by the designed encryption scheme. Besides, the encryption performance is given in detailed. The experimental results show that the proposed encryption scheme has quite good encryption performance. Therefore, it can ensure the information security of SG.

**Keywords**: Smart Grid, Chaotic System, Data Encryption, Memristor, Information Security

## 1 INTRODUCTION

The SG is a system based on communication and information technology in the generation, delivery, and consumption of energy power. It (Kimani et al. (2019); Ferrag et al. (2018)) begins to involve application areas such as smart factory, traffic network and gas system (as shown in Figure 1).

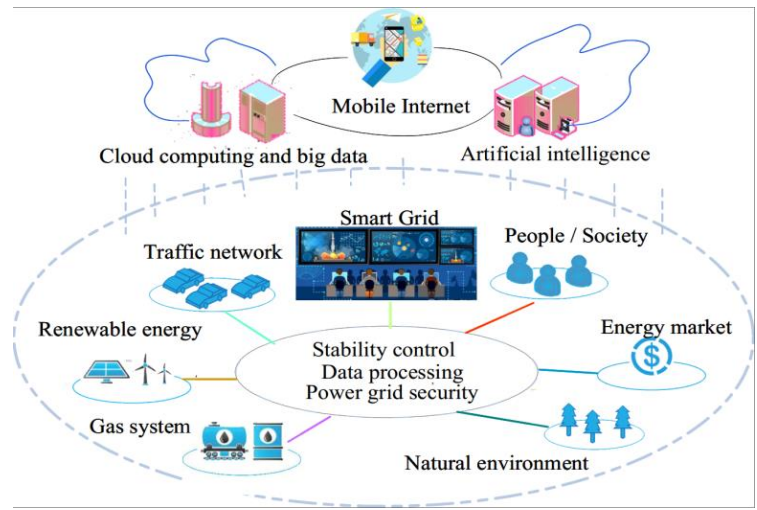

**Figure 1**. The Block diagram of distributed power energy system

The SG has great openness and interconnection, and there may be some potential problems in information collection and communication. The control of all infrastructure in SG relies on the internet. Therefore, the information security and privacy preservation in the SG is extraordinarily important (Li et al. (2022a)). If the SG is maliciously attacked, the fact that data loss and tampering may happen (Li et al. (2019)). It will seriously affect the normal operation of the SG, which will even lead to the consequences of system instability (Li et al. (2015, 2022d)). Then the safe and smooth operation of the SG is challenging to realize. As an example, remote data acquisition systems are usually installed in SG, which can be accessed without authorization and passwords. These devices are easily attacked or controlled by illegal users. Once the infrastructure has been maliciously controlled in unsupervised situation and it may bring huge economic losses. In addition, due to the inherent uncertainty of renewable energy (Li et al. (2022b,c)), the safe operation of power systems with high-penetration renewables is facing greater challenges.

In sharp contrast to the important position of SG, the attention to its network and information security are still insufficient, which is also the reason for the frequent occurrence of power system accidents. Therefore, in the SG, designing an encryption algorithm to achieve information security is essential.

Many scholars have developed numerous researchs for the information security of SG. In May 2021, on the basis of the homomorphic encryption, Zhao et al. (2021) proposed a data aggregation and realtime electricity price billing scheme to reduce the computing cost. In July 2021, Singh et al. (2021) proposed a data aggregation model on the basis of the deep learning and homomorphic encryption. In May 2021, based on the partially homomorphic encryption (PHE), Wu et al. (2021) introduced a privacy-preserving distributed optimal power flow (OPF) algorithm. In Jan 2022, Hussain et al. (2021) preserved the privacy of customers by the homomorphic encryption in the SG. Even though academics have studied excellent approaches for the information security of SG, there yet be two issues to be handled:

(1) The homomorphic encryption contains a certain amount of operations, which is difficulty to be implemented.

(2) For the sake of guaranteeing the infor2mation security, there exist a heavy computation burden caused by the homomorphic encryption.

Since the mid-1990s, many scholars have found that there is a close relationship between chaotic system and cryptography. A chaotic system has a series of features, such as sensitivity to the initial value, system parameters, ergodicity, unpredictability of orbit, and good pseudo-randomness. These characteristics can just meet the requirements of encryption. Therefore, chaos has been extensively applied in numerous realms, such as chaos control (Tian et al. (2021); Li et al. (2020)), chaotic spread spectrum communication (Yuan et al. (2021); Xiao et al. (2018)), secure communication (Zhao et al. (2020); He et al. (2020)), chaos optimization (Shi et al. (2008))and so on. Besides, the application of chaos in cryptography is not difficulty to be realized. The algorithm exhibits great performances with fast encryption speed and large key space.

These advantages make the algorithm suited for encrypting a lot of data. Then it extremely simplifies the design of traditional sequence cipher. Therefore, chaos has unique superiority in the realm of encryption and broad development prospects.

To address the issues as mentioned above, a data encryption algorithm combined with the chaotic sequence is introduced. Then an unusual 7D-CCS with cubic memristor is put forward to create pseudo-random sequences. The 7D-CCS has complex dynamic characteristics and can generate pseudorandom sequences with high pseudorandomness. The originality and contributions of this paper are summed up as follows.

(1) The designed algorithm only includes scrambling and diffusion operation. It is easy to be implemented.

(2) Based on the cubic memristor, the 7D-CCS is proposed to create pseudo-random sequences with good pseudorandomness to ensure the information security. Besides, the 7D-CCS is easy to generate key sequence.

The rest of this paper is organized as follows: In Section 2, the features of the 7D-CCS are introduced. In Section 3, design an encryption scheme. Besides, it is compared with that of others by some performance indexes. In Section 4, the data and monitoring images in SG are processed by the proposed encryption scheme, and the security analysis are provided. Conclusions are given in Section 5.

**2 THE 7D-CCS WITH CUBIC MEMRISTOR**

The mathematical expression (Yang et al. (2019)) of cubic nonlinear memristor is

$$q(\varphi) = a\varphi + b\varphi^3 (a, b > 0) \tag{1}$$

where $a$, $b$ are the positive constants. $\varphi$ is an independent variable.

Then the derivative of memristor $W(\varphi)$ is defined by

$$W(\varphi) = dq(\varphi)/d\varphi = a + 3b\varphi^2 \tag{2}$$

The real chaotic system with the cubic nonlinear memristor is given by

$$\begin{cases} \dfrac{dx}{dt} = \alpha(y - x + dx - W(\varphi)x) \\ \dfrac{dy}{dt} = x - y + z \\ \dfrac{dz}{dt} = rz - \beta y \\ \dfrac{d\varphi}{dt} = x \end{cases} \tag{3}$$

where $\alpha, \beta, r, d$ are positive constants. $x$, $y$, $z$ are independent variables.

System (3) is extended to the complex field, where $x = x_1 + jx_2$, $y = x_3 + jx_4$, $z = x_5 + jx_6$ and $\varphi = x_7$. $x_i(i = 1, ..., 7)$ are independent variables. $j$ is the imaginary number.

In system (3), the real and imaginary parts are divided. We can get:

$$\begin{cases} \dfrac{dx_1}{dt} = \alpha(x_3 - x_1 + dx_1 - W(x_7)x_1) \\ \dfrac{dx_2}{dt} = \alpha(x_4 - x_2 + dx_2 - W(x_7)x_2) \\ \dfrac{dx_3}{dt} = x_1 - x_3 + x_5 \\ \dfrac{dx_4}{dt} = x_2 - x_4 + x_6 \\ \dfrac{dx_5}{dt} = rx_5 - \beta x_3 \\ \dfrac{dx_6}{dt} = rx_6 - \beta x_4 \\ \dfrac{dx_7}{dt} = x_1 \end{cases} \quad (4)$$

Finally, through the chao attractor, Lyapunov exponents, bifurcation diagram, 0-1 test, and complexity analysis, we discuss the dynamic features of the system (4).

## 2.1 Chaos Attractor

Set $\alpha = 10$, $x = \dfrac{-b \pm \sqrt{b^2 - 4ac}}{2a}$, $\beta = \dfrac{100}{7}$, $r = 0.1$, $d = \dfrac{9}{7}$, $a = \dfrac{1}{7}$, $b = \dfrac{2}{7}$. For initial conditions (0.1, 0.1, 0.1, 0.1, 0.1, 0.1, 0.1), the attractor of 7D-CCS are showed in Figure 2.

The Lyapunov exponent (Sutter et al. (2021)), one of the numerical features, is used to recognize chaotic motion quantitatively. If the motion in this direction is stable, the value is negative. If the motion in this direction is unstable, the value is positive. If Lyapunov exponents include positive, negative values and zero, the system is chaotic. In the system (4), the Lyapunov exponents are LE1= 2.041, LE2 = 0.425, LE3 = 0.189, LE4 =0, LE5 = -0.041, LE6 = -3.355, LE7 = -4.205. The Lyapunov exponents are showed in Figure 3. The Lyapunov exponent of the 7D-CCS is (+, +, +, 0, -, -, -). Hence, the 7D-CCS is chaotic.

## 2.2 Bifurcation Diagram of the 7D-CCS

The bifurcation diagram (Marszalek and Sadecki (2019)) could distinctly show the complete process of the nonlinear system into chaos. In the bifurcation diagram, if there exist a large number of point of density caused by the infinite bifurcation, it indicates that the system is chaotic. In Figure 4, the bifurcation diagram of the 7D-CCS is shown. As shown in the Figure 4, with the change of $\alpha$, the system continually forks among different states. Finally, the system (4) comes to a chaotic state.

## 2.3 0-1 test

The 0-1 test Karimov et al. (2021) is a method which directly calculate $p(n)$ and $q(n)$ to judge the state of nonlinear system. The 0-1 test method is as follows:

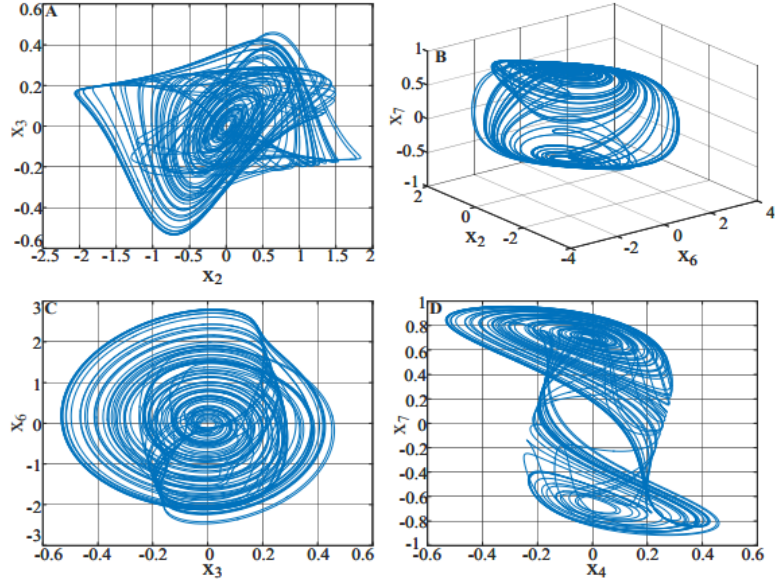

**Figure 2.** Phase portraits of system (4), (A) $x_2 - x_3$ (B) $x_6 - x_2 - x_7$ (C) $x_3 - x_6$ (D) $x_4 - x_7$

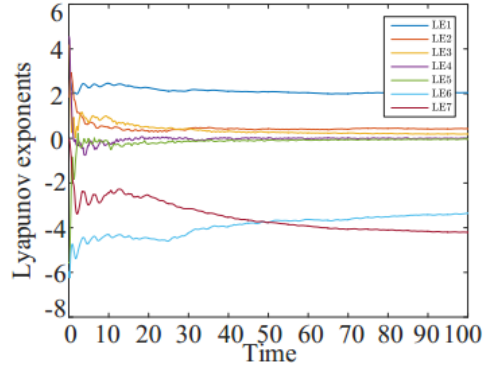

**Figure 3.** The Lyapunov exponent curves of 7D-CCS

**Step 1**: Let $X(k)(k = 1, 2, ..., N)$ be a test sequence.

**Step 2**: Calculate the sequence $p(n)$ and $q(n)$:

$$p(n) = \sum_{j=1}^{n} X(k)\cos(kc), k = 1, 2, ..., N \qquad (5)$$

$$q(n) = \sum_{j=1}^{n} X(k)\sin(kc), k = 1, 2, ..., N \qquad (6)$$

where $c \in (0, \pi)$.

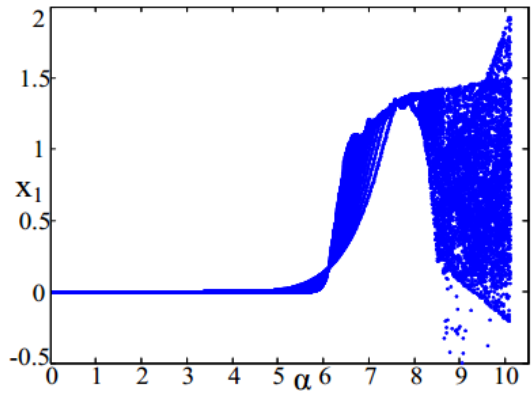

**Figure 4.** The bifurcation diagram of 7D-CCS

If the trajectory diagram of $p(n)$-$q(n)$ is represented by the Brownian motion, system is in a chaotic state The "0-1 test" diagram of 7D-CCS is exhibited in Figure 5. Then the Brownian motion can be seen. Hence, the system (4) is chaotic.

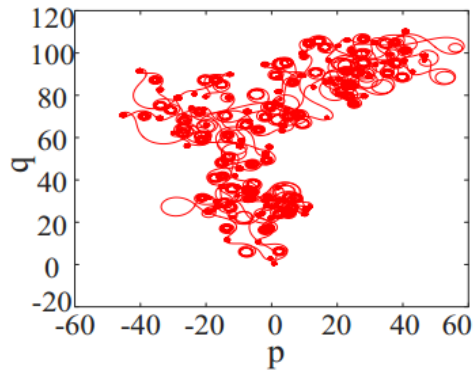

**Figure 5.** The 0-1 test diagram of 7D-CCS

## 2.4 Complexity Analysis of 7D-CCS

The $SE$ (Yu et al. (2020)) and $C_0$ (Chen et al. (2020)) algorithm, based on Fourier transform and wavelet transform, are spectral entropy algorithm until now. When the two parameters vary, the chromatogram is introduced to verify and analyze the complexity. The chromatogram of 7D-CCS is exhibited in Figure 6. The lighter the hue is, the lower the complexity is.

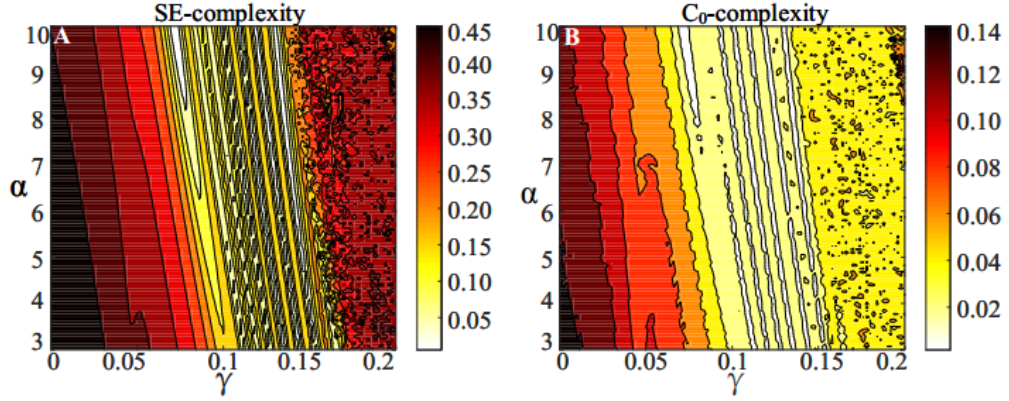

**Figure 6.** The chromatogram, (A) $x_1$ sequence chromatogram by $SE$ algorithm (B) $x_1$ sequence chromatogram by $C_0$ algorithm

Based on these performance metrics, the nonlinear dynamic features of 7D-CCS are discussed. Set $\alpha = 10, d = 9/7, \beta = 100/7, r = 0.1, a = 1/7$, and $b = 2/7$. Then the pseudo random sequences are created by the 7D-CCS. They can meet the requirements of the designed algorithm.

## 3 THE DESIGNED ALGORITHM AND ITS DISCUSSION

Based on the system (4), a novel data encryption algorithm is introduced. The designed algorithm is exhibited in Figure 7. The procedures of algorithm are as follows:

**Step1**: The experiment environment is Intel (R) Core (TM) i5-9300H CPU @ 2.40GHz, and the random-access memory (RAM) adopted is 8 GB. The R, G and B channel are get by separating the image channel.

**Step2**: Use random function to randomly transform the position of the three primary color pixel value. Call them R1,G1 and B1.

**Step3**: Transform the position of the three primary color pixel value according to Arnold transform. Name them R2,G2 and B2. Arnold transform is as follows:

$$\begin{pmatrix} xx_{n+1} \\ yy_{n+1} \end{pmatrix} = \begin{pmatrix} 1 & A \\ B & AB+1 \end{pmatrix} \begin{pmatrix} xx_n \\ yy_n \end{pmatrix} \mod \begin{pmatrix} M \\ N \end{pmatrix} + \begin{pmatrix} 1 \\ 1 \end{pmatrix} \quad (7)$$

where $M$ and $N$ are the row and column of the image matrix. The pseudo-random matrices $A$ and $B$ with the sizes of $M \times N$ are generated from the proposed chaotic sequence. Let the coordinates of pixels in digital image be $xx_n, yy_n \in [0, 255]$.

**Step4**: XOR the generated seven dimensional pseudo-random sequence with R2, G2 and B2 image seven times, and the order of XOR is random.

Set $\alpha = 10, \beta = \frac{100}{7}, r = 0.1, d = \frac{9}{7}, a = \frac{1}{7}, b = \frac{2}{7}$ in system 4, and initial condition is (0.1, 0.1, 0.1, 0.1, 0.1, 0.1, 0.1). Standardized test "Lena" in the size of 256×256 is selected in the above algorithm. The encryption process is shown in Figure 8. Figure 8(A) is the standard test picture Lena, Figure 8(B) is the scrambled picture, Figure 8(C) is the encrypted picture and Figure 8(D) is the decrypted picture. Figure 8(C) conceals the characters of the original image and a malicious third party cannot be directly identified.

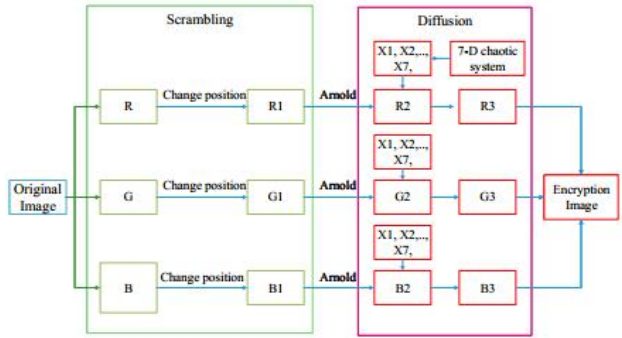

**Figure 7.** The image algorithm flow chart

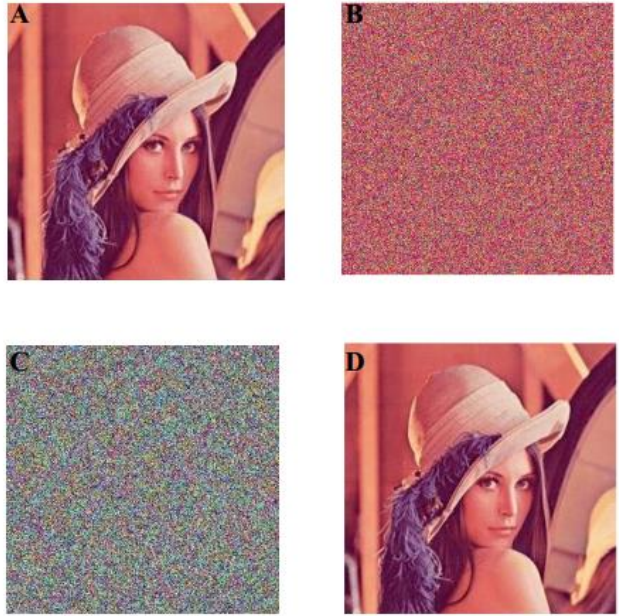

**Figure 8.** The diagram of experimental results, (A) Standard test picture Lena (B) Scrambled picture (C) Encrypted picture (D) Decrypted picture

## 3.1 Reconstruction Quality Analysis

The peak signal-to-noise ratio ($PSNR$) is introduced to investigate the visual quality of the reconstructed image. When the $PSNR$ is greater than 30dB but less than 40dB, the distortion of image is small. The $PSNR$ method is as follows:

**Step 1**: Calculate the mean square deviation ($MSE$):

$$MSE(f,g) = \frac{1}{MN} \sum_{i=1}^{M} \sum_{j=1}^{N} (f_{ij} - g_{ij})^2 \tag{8}$$

**Step 2**: Calculate $PSNR$:

$$PSNR(f,g) = 20\log_{10}(255/MSE(f,g)) \tag{9}$$

where $f, g$ are the the pixel values of original image and decrypted image. $M$ and $N$ represent the row and column of the image matrix, respectively. Calculate the $PSNR$ between Figure 8(a) and Figure 8(d) and it is approximately 30dB. The distortion of Figure 8(a) and Figure 8(a) is small.

The structural similarity ($SSIM$) is another quota to estimate the similarity of two images. The formula is as follows:

$$\begin{cases} L(X,Y) = \dfrac{2u_X u_Y + C_1}{u_X^2 + u_Y^2 + C_1} \\ C(X,Y) = \dfrac{2\sigma_X \sigma_Y + C_2}{\sigma_X^2 + \sigma_Y^2 + C_2} \\ S(X,Y) = \dfrac{\sigma_{XY} + C_3}{\sigma_X \sigma_Y + C_3} \\ SSIM = L(X,Y)C(X,Y)S(X,Y) \end{cases} \tag{10}$$

where $u_x$ is the the mean of image $X$. $u_y$ is the the mean of image $Y$. $\sigma_X$ is the variance of image $X$. $\sigma_Y$ is the variance of image $Y$. $\sigma_{XY}$ is the covariance of images $X$ and $Y$. In order to avoid instability, when denominator is up to zero, $C_1$ and $C_2$ are two constants with small value. $C_3 = \frac{1}{2}C_1$. $L(X,Y)$ is the luminance, $C(X,Y)$ denotes the contrast, $S(X,Y)$ represents the structure.

The scope of *SSIM* is [0, 1]. When the value approaches to 1, it represents the good resemblance between the two images When the value approaches to 0, it indicates less resemblance. Calculate the $SSIM$ between Figure 8(A) and Figure 8(D) and it is 1. The calculation results show that the structure of the Figure 8(D) is the same as that of the Figure 8(A).

### 3.2 Correlation Coefficient

In order to prevent the original information from being cracked through the similarity between pixels, it is very necessary to remove the correlation between adjacent pixels. Firstly, choose $N$ pairs of pixels in the original image randomly. Then noted them as $(u_i, v_i)$, $i \in [1, N]$. The formula of the correlation coefficient is shown as follows:

$$E(u) = \frac{1}{N} \sum_{i=1}^{N} u_i \tag{11}$$

$$D(u) = \frac{1}{N} \sum_{i=1}^{N} (u_i - E(u))^2 \tag{12}$$

$$Cov(u,v) = \frac{1}{N}\sum_{i=1}^{N}(u_i - E(u))(v_i - E(v)) \qquad (13)$$

$$r_{uv} = \frac{Cov(u,v)}{\sqrt{D(u)} * \sqrt{D(v)}} \qquad (14)$$

where $E$ is the mean value of pixel. $D$ is the variance of pixels. $Cov$ is the covariance of pixels $r_{uv}$ is the correlation coefficient.

In plaintext and ciphertext images, 8000 pairs of adjacent pixel values are haphazardly choose from the horizontal, vertical and diagonal directions. The correlation coefficient between two adjacent pixels is calculated. The value of the correlation coefficient of adjacent pixels is from -1 to 1. If the value is approach to 1, the correlation is high. Correspondingly, the adjacent pixels are basically uncorrelated if the value is close to -1. From the Table 1, we can know that the correlation coefficient of the designed algorithm is lower than that of other algorithms.

It shows that the designed algorithm can almost break the correlation between pixels.

Table 1. ADJACENT PIXELS CORRELATION COMPARISON

| Picture | Horizontal | Vertical | Diagonal |
|---|---|---|---|
| Lena | +0.9817 | +0.9603 | +0.9483 |
| Cipher Lena (proposed method) | -0.0177 | -0.0321 | -0.0026 |
| Lena(Zhou et al. (2020)) | +0.0083 | -0.0054 | -0.0010 |
| Lena(Hosseinzadeh et al. (2019)) | -0.0003 | +0.0012 | -0.0017 |
| Lena(Pak et al. (2019)) | -0.0024 | +0.0035 | +0.0014 |
| Lena(Huang et al. (2018)) | +0.0010 | -0.0031 | -0.0008 |

### 3.3 Histogram

In digital image, the distribution of each gray level can be counted by the histogram. The Figure 9 shows that the pixel distribution of each pixel level of the three primary color matrixes. Figure 9(A), Figure 9(B) and Figure 9(C) show fluctuates greatly, and the peak and trough values differ extremely. The frequency of some pixel values is large, while that of others is very small. After encryption, the pixel distribution of each pixel level of the three primary color matrixes is relatively uniform, and the value frequency of each pixel value is basically the same, which well conceals the distribution law of the original image.

### 3.4 Information Entropy

To the pixel values, the mean uncertainty can be reflected by the information entropy. The formula is exhibited as follows:

$$H = -\sum_{i=1}^{255} p(x_i)\log_2 p(x_i) \qquad (15)$$

where $p(x_i)$ is the probability of gray value. The larger the image information entropy is (the maximum value is 8), the more equivalent the distribution of pixels is. The nonrandom distribution of image pixels

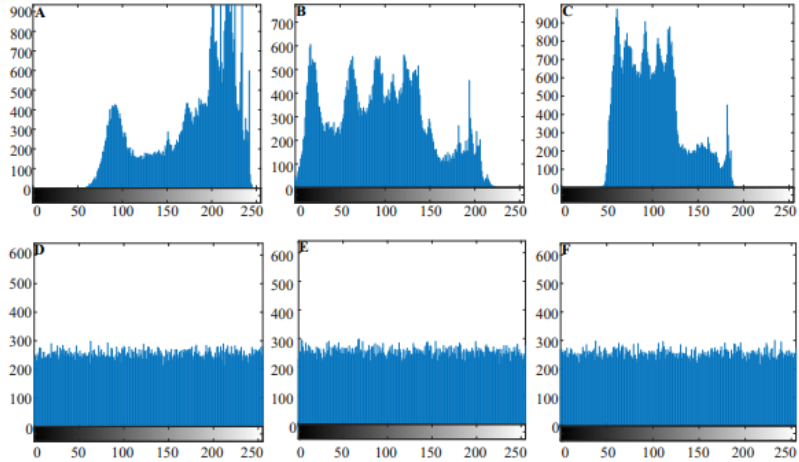

**Figure 9.** The histogram of "Lena", (A) R channel in plain image (B) G channel in plain image (C) B channel in plain image (D) R channel in cipher image (E) G channel in cipher image (F) B channel in cipher image

indicates that the encryption effect is quite good. As shown in Table 2, in this algorithm, compared with other algorithms, the information entropy of encrypted image is more approach to 8. We could know that the proposed method has enough ability to withstand differential attacks.

**Table 2.** THE INFORMATION ENTROPY OF PICTURE

| Picture | Original Picture | Encrypted Picture |
|---|---|---|
| Lena | 7.4375 | 7.9991 |
| Lena(Zhou et al. (2020)) | 7.4375 | 7.9972 |
| Lena(Hosseinzadeh et al. (2019)) | 7.4375 | 7.9971 |
| Lena(Pak et al. (2019)) | 7.4375 | 7.9972 |
| Lena(Kumar Patro and Acharya (2019)) | 7.4375 | 7.9971 |

## 3.5 Security Key Space

Assume the accuracy of the computer memory is $10^{15}$, then the size of the key space of each key is $10^{15}$. There are 7 variable values and 6 system parameter values in system (4) and the key space can reach $\left(10^{15}\right)^{13} = 10^{195} \approx 2^{650}$. Hence, the key space of the designed algorithm is greatly ample. With a sufficient security level, the algorithm is able to resist differential cryptanalysis.

## 4 THE ENCRYPTION SCHEME OF SMART GRID

### 4.1 The Simulation of Image Encryption

In the SG, in order to confirm the operation status of the equipment, the remote data acquisition system should transmit the monitoring image to the control center. When the control center finds the equipment failure, it will shut down the equipment for maintenance.

However, in the remote data acquisition system, the monitoring images are easy to be obtained by illegal users because it do not need authorization and password. When the equipment is in normal operation, the illegal user transmits the monitoring image of equipment failure to the control center, resulting in the shutdown of the equipment. Then it will bring huge economic losses.

Therefore, in the remote data acquisition system, it is of practical significance to encrypt the monitoring image in real time, and they can be encrypted and transmitted immediately. Then the illegal user can not obtain the monitoring image.

The "Picture 1" and "Picture 2" transmitted in SG will be encrypted by using the above algorithm. The size of "Picture 1" is 660 × 783. The size of "Picture 2" is 456 × 639. The encryption process of "Picture 1" is exhibited in Figure 10. Figure 10(A) is the original "Picture 1". Figure 10(B) is the scrambled "Picture 1". Figure 10(C) is the encrypted "Picture 1". The characters in the original image cannot be identified directly from this image and Figure 10(D) is the decrypted "Picture 1".

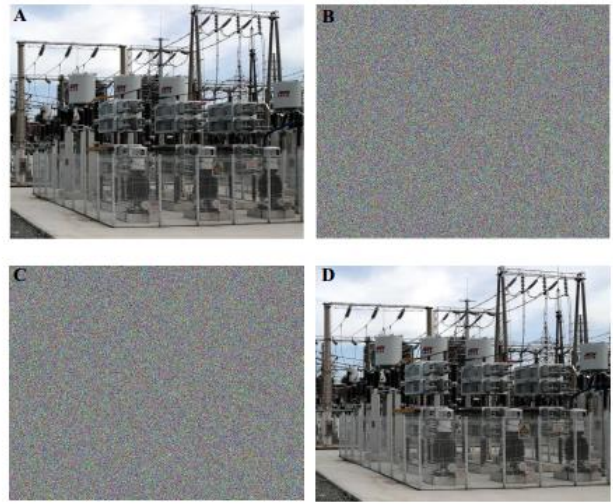

**Figure 10.** The diagram of experimental results, (A) Electronic equipment Picture 1 (B) Scrambled Picture 1 (C) Encrypted Picture 1 (D) Decrypted Picture 1

The encryption of "picture 2" is similar to that of "picture 1", which is shown in the Figure 11.

Then the metrics mentioned in section 3 are used to analyze them.

### 4.1.1 Histogram

According to the Figure 12 and Figure 13, after encryption, the pixel distribution of each pixel level of the three primary color matrices is relatively uniform.

### 4.1.2 Correlation Coefficient

Compare the cipher image with the original image, the adjacent pixels have almost no correlation, even negative correlation. That can be known in the Table 3.

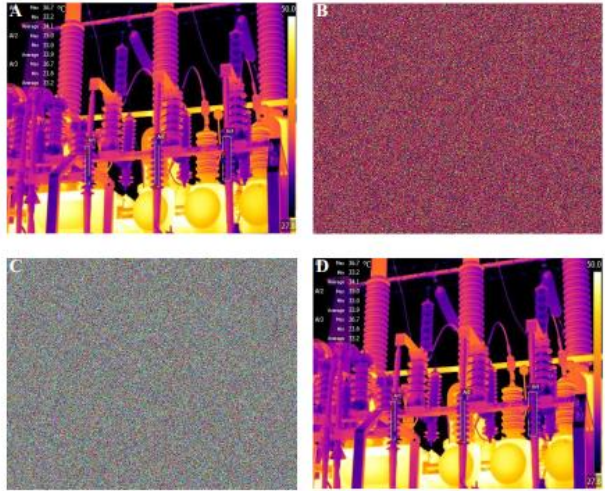

**Figure 11.** The diagram of experimental results, (A) Electronic equipment Picture 2 (B) Scrambled Picture 2 (C) Encrypted Picture 2 (D) Decrypted Picture 2

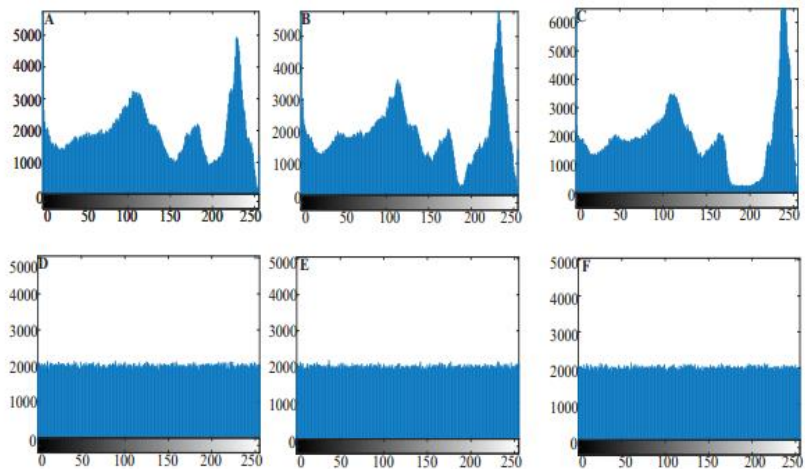

**Figure 12.** The histogram of the "picture 1", (A) R channel in plain picture 1 (B) G channel in plain picture 1 (C) B channel in plain picture 1 (D) R channel in cipher picture 1 (E) G channel in cipher picture 1 (F) B channel in cipher picture 1

### 4.1.3 Information Entropy

In the Table 4, the information entropy is close to 8 (max is 8). The results show that encrypted images are resistant to differential cryptanalysis.

### 4.1.4 Sensitivity of key

The initial conditions is (0.10001, 0.10001, 0.10001, 0.10001, 0.10001, 0.10001, 0.10001). In system (3), the value of system parameters remain unchanged. The cipher picture 1 and 2 are decrypted by the generated key which is in the above initial conditions. The Figure 14 is the decryption result of the wrong key. According to the Figure 14, with small change of key, it cannot successfully decrypt the picture 1 and 2. We could know that the proposed algorithm has great sensitivity to the key.

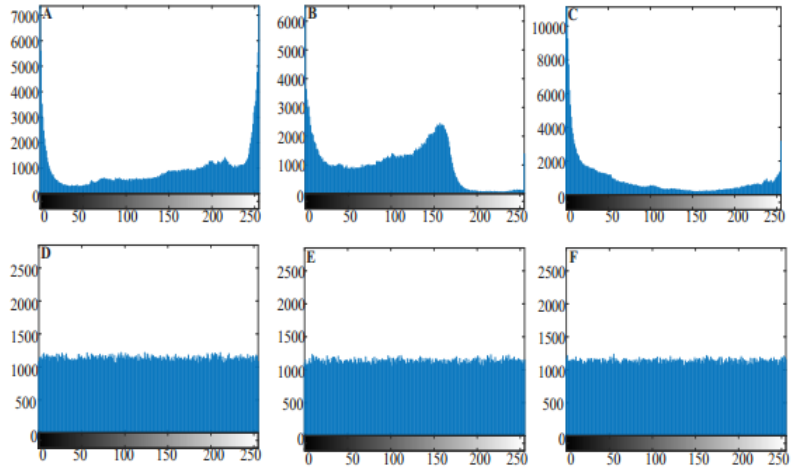

**Figure 13.** The histogram of the "picture 2", (A) R channel in plain picture 2 (B) G channel in plain picture 2 (C) B channel in plain picture 2 (D) R channel in cipher picture 2 (E) G channel in cipher picture 2 (F) B channel in cipher picture 2

**Table 3.** ADJACENT PIXELS CORRELATION OF PICTURES

| Picture | Horizontal | Vertical | Diagonal |
|---|---|---|---|
| Picture 1 | +0.9108 | +0.9122 | +0.8530 |
| Picture 2 | +0.9722 | +0.9580 | +0.9374 |
| Cipher Picture 1 | -0.0223 | -0.0026 | +0.0040 |
| Cipher Picture 2 | +0.0084 | -0.0205 | -0.0014 |

**Table 4.** THE INFORMATION ENTROPY OF PICTURE

| Picture | Original Picture | Encryption Picture |
|---|---|---|
| Picture 1 | 7.6158 | 7.9999 |
| Picture 2 | 7.1023 | 7.9998 |

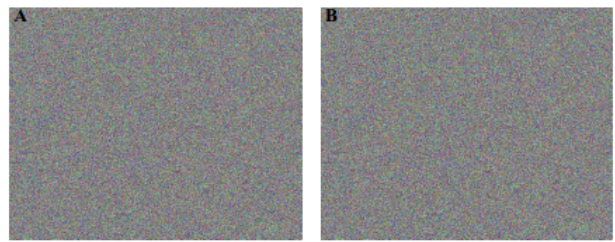

**Figure 14.** The results of wrong key, (A) Wrong key with picture 1 (B) Wrong key with picture 2

**Table 5.** RECONSTRUCTION QUALITY ANALYSIS

| Picture | PSNR(dB) | SSIM |
|---|---|---|
| Figure 11(a) and (d) | 27.49 | 1 |
| Figure 12(a) and (d) | 26.89 | 1 |

### 4.1.5 Reconstructed Image Quality

According to the Table 5, the quality loss of reconstructed images is relatively small. They are the same as the original images.

## 4.2 Data Encryption

The dataset used to check the security of above algorithm is the temperature data in the SG. The Modbus Protocol is used to transmit temperature data. In the Modbus protocol, the data is as follows: $0x13\ 0x04\ 0x00\ 0x00\ 0x00\ 0x01\ 0x32\ 0xB8$ where is 8 bytes and hexadecimal.

The flow chart of data encryption algorithm is shown in Figure 15. The steps are as follows

**Step1**: Use random function to randomly change the position of the temperature data. Name it Data 1.

**Step2**: Change the position of Data 1 according to Arnold transform. Name it Data 2.

**Step3**: XOR the generated seven dimensional pseudo-random sequence with Data 2 seven times, and the order of XOR is random.

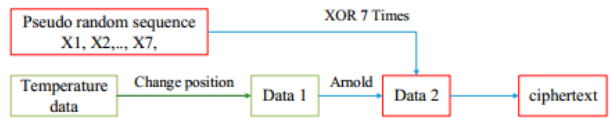

**Figure 15.** The data algorithm flow chart

The ciphertext data composition is as follows: $0xFC\ 0xCF\ 0xFC\ 0xFC\ 0xFC\ 0xFC\ 0xF8\ 0xF8$ where is 8 bytes and hexadecimal.

After decryption, the data composition is the same as the initial data. The security of cipher text mainly depends on whether the key is random. If the key is random and variable, the security of ciphertext can be guaranteed. Next, from the NIST test to analyze the randomness of the key.

The NIST statistical test suit is composed of 15 statistical tests, which can detect the randomness of the sequences created by the 7D-CCS. Generally speaking, the statistical test is successful when the test result is between 0.01 and 1. Bsides, the test sequence has excellent randomness if the value is large. For simplicity, 20, 000, 000 real numbers, created by the 7D-CCS, are adopted as the test data in the NIST test. The NIST test results are shown in Table 6, which are between 0.01 and 1. It means that the statistical tests are successful. Then it also verified that the key has quite good randomness.

**Table 6.** NIST TEST

| Test category | Value | the test |
|---|---|---|
| Approximate Entropy | 0.437274 | pass |
| Block Frequency | 0.122325 | pass |
| Cumulative Sums | 0.834308 | pass |
| FFT | 0.437274 | pass |
| Frequency | 0.739918 | pass |
| Linear Complexity | 0.213309 | pass |
| Longest Run | 0.122325 | pass |
| NonOverlapping Template | 0.991468 | pass |
| Overlapping Template | 0.213309 | pass |
| Random Excursions | 0.964295 | pass |
| Random Excurions Variant | 0.710216 | pass |
| Rank | 0.122325 | pass |
| Runs | 0.911413 | pass |
| Serial | 0.834308 | pass |
| Uiversal | 0.350485 | pass |

## 5 CONCLUSIONS

Based on the 7D-CCS, a data process scheme is introduced. First of all, the 7D-CCS with memristor is proposed, which is derived from the real 4D chaotic system with the cubic nonlinear memristor. Secondly, the dynamic characteristics are analyzed by some performance indexes. Thirdly, the standard test image Lena is selected as encrypted object. Then compare it with others. Finally, the temperature data and monitoring images are encrypted in the verification experiments.

Simulations include *PSNR*, *SSIM*, histogram analysis, information entropy analysis, correlation analysis, sensitivity of key analysis, key space analysis and NIST test.

The experimental results indicate that the designed algorithm has excellent security performance. Therefore, the designed algorithm is suitable for SG in which the high security is required.

Besides, there are many tasks needed to be further studied.

(1) At present, only the security of the designed algorithm is analyzed. The field experiments need to be carried out.

(2) In the research of SG, the efficiency of data encryption is also critical. The encryption time should be considered in the future research work.

## AUTHOR CONTRIBUTIONS

Conceptualization, L. Kou and Z. Huang; methodology, L. Kou, Z.Huang and C. Jiang; software, Z. Huang; validation, Z. Huang and C. Jiang; formal analysis, F. Zhang and Z. Huang; investigation, H. Li and H. Liu; resources, W. Ke; data curation, W. Ke and J. Wan.

## FUNDING

This work is funded by Major scientific and technological innovation projects of Shandong Province (No.2019JZZY010731 and No.2020CXGC010901), the project of "Youth Innovation and technology support plan" for colleges and universities in Shandong Province (2021KJ025), International Collaborative Research Project of Qilu University of Technology (No.QLUTGJHZ2018020).